\documentclass[12pt]{article}
\usepackage{graphicx}

\usepackage{amssymb}
\usepackage{amsmath}
\usepackage{amsfonts}

\usepackage{dcolumn}

\def\ni{\noindent}

\def\xu{\hat{\mathbf x}}

\def\zu{\hat{\mathbf z}}
\def\be{\begin{equation}}
\def\ee{\end{equation}}
\def\pv{\mathbf p}
\def\rv{\mathbf r}
\def\vv{\mathbf v}
\def\Av{\mathbf A}
\def\Av{\mathbf A}
\def\Ev{\mathbf E}
\def\Fv{\mathbf F}
\def\Lv{\mathbf L}
\def\Mv{\mathbf M}

\def\vT{\vv_{\rm T}}

\def\en{{\cal E}}
\def\ER{\epsilon_{\rm R}}  
\def\Re{{\rm Re\,}}
\def\Im{{\rm Im\,}}

\def\os{\Phi}
\def\fxi{\chi}
\def\feta{\Theta}
\def\profil{\hat\psi}
\def\exi{\lambda}
\def\eeta{\mu}

\def\np{j}
\def\nm{k}
\def\NP{J}
\def\NM{K}
\def\nA{n_{\rm A}}
\def\te{t^*}   
\def\vLF{$[\mathbf{v},\mathbf{L},\mathbf{F}]$ }
\def\crt{_{\rm cr}}

\begin{document}

\title{
Transverse momentum asymmetry of the extracted electron in field ionization of a hydrogen atom with angular momentum} 


\author{X. Artru$^{1,*}$, E. Redouane-Salah$^{2,3}$}

 
%


%

\maketitle

 \begin{center}
$^1$ {Universit\'e de Lyon;  CNRS/IN2P3; Universit\'e Lyon 1; Institut de Physique Nucl\'eaire de Lyon, 69622 Villeurbanne}, France. 
\\
$^2$ Ernest Orlando Lawrence Berkeley National Laboratory, 
University of California, Berkeley, CA 94720, USA \\
$^3$ Universit\'e de M'sila, D\'epartement de Physique, Algeria \\
and Laboratoire de physique math\'ematique et physique subatomique, Universit\'e de Constantine 1, Algeria 
\end{center}

\begin{abstract}
The tunneling ionization of a hydrogen atom excited in the presence of a static electric field is investigated
for the case where, before being extracted, the electron has an orbital angular momentum $\Lv$ perpendicular to the field $\Ev$. The escaping electron has a nonzero mean transverse velocity $\langle\vT\rangle$ in the direction of $\Ev\times\langle\Lv\rangle$.
This asymmetry is similar to the Collins effect in the fragmentation into hadrons of a transversely polarized quark.  
In addition, the linear Stark effect make $\langle \Lv \rangle$ and $\langle\vv_{\rm T}\rangle$ oscillate in time. The degree of asymmetry is calculated at leading order in $\Ev$ for an initial state of maximum transverse $\langle\Lv\rangle$. 
The conditions for the observation of this asymmetry are discussed. 
\end{abstract}


\ni keywords  :    {field ionization ; electron imaging ; Stark effect ; tunnel effect ; Gamow states ; Runge-Lenz vector ; Collins effect}

\ni PACS numbers: 32.60.+i , 13.88.+e

\vskip 1cm

\ni $^*$ email: {x.artru@ipnl.in2p3.fr} 

\begin{figure}   [t]
\centering
\begin{tabular}{cc}
\includegraphics*[bb= 80 400 650 750, width=70mm]{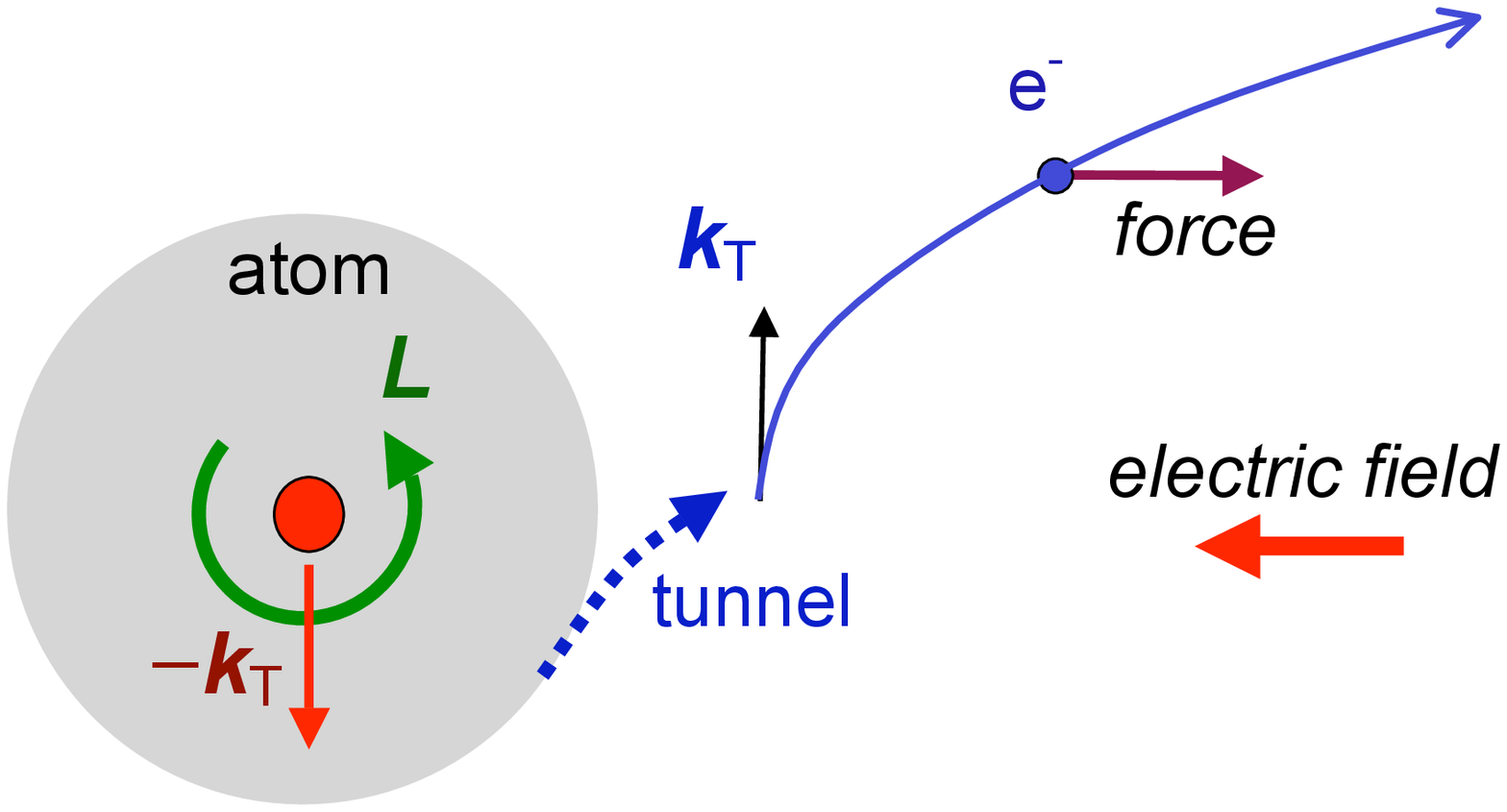} 
&  
\includegraphics*[bb= 70 530 550 800, width=70mm]{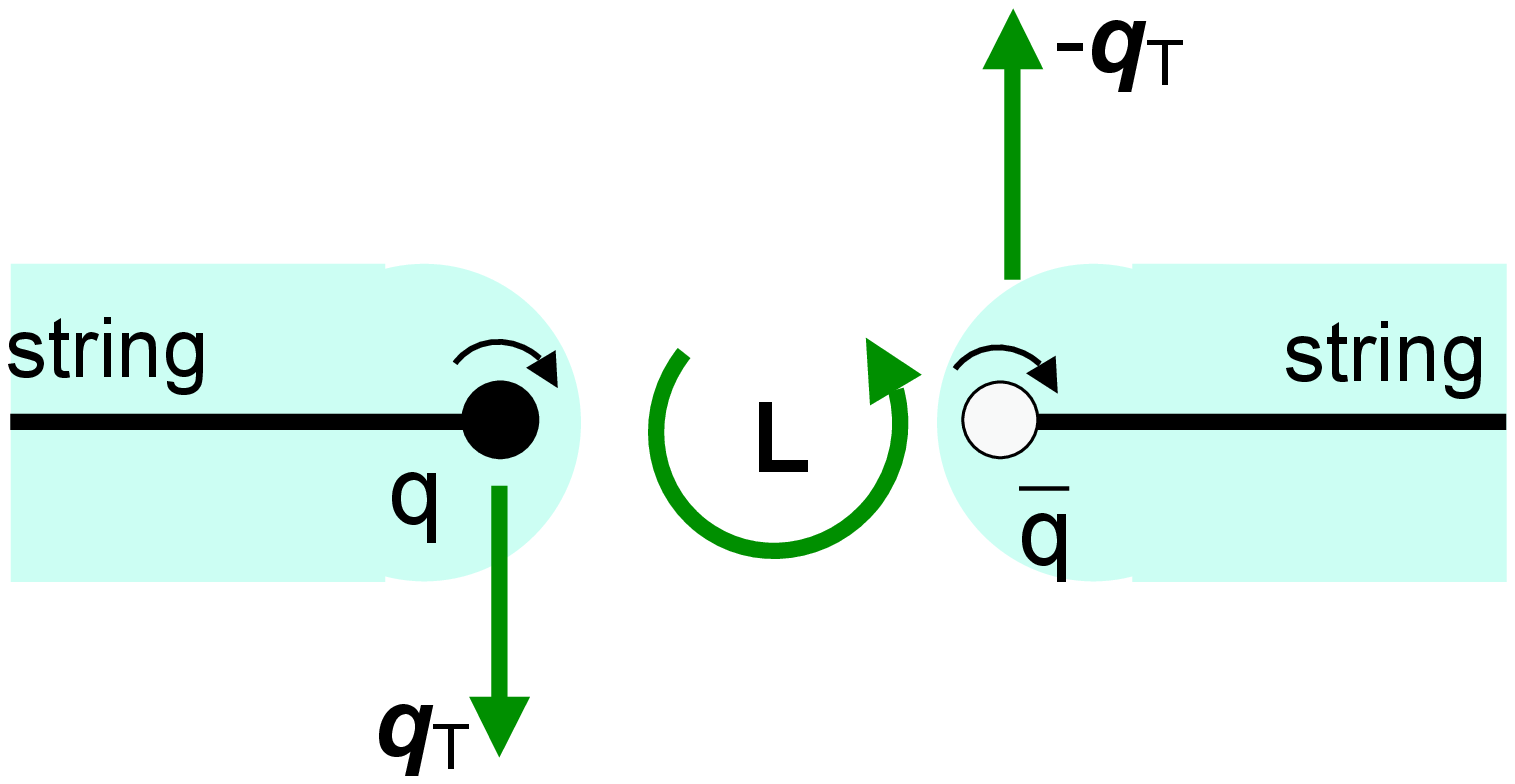}
\end{tabular}
\caption{\footnotesize
Left: semi-classical motion of the electron extracted from the hydrogen atom by a strong field ${\bf E}$, when the electron is initially in a $L_y = +1$ state. 
Right:  
String and $^3{\rm P}_0$ mechanism correlating the transverse momentum and the transverse polarization of a quark created in string decay \cite{ARTRU-CZ-Y,ARTRU-CZ}. }
\end{figure}

\section{Introduction}

An atom placed in a strong static electric field $\mathbf{E}$ may be ionized through the tunneling effect if the initial electron energy is below the saddle point of the sum of the atomic and external potentials. The calculation, to lowest order in $\Ev$, of the ionization rate $\gamma$ for a hydrogen atom in the ground state ($n=1$) is given in textbooks \cite{LANDAU}. A large amount of work has been devoted to the generalization to $n\ge2$, with the inclusion of higher orders in $\Ev$; see \cite{YAMABE} and references therein for analytical calculations, \cite{DAMBURG,LUC,KOLOSOV}  and references therein for numerical calculations.  

The distribution in transverse velocity $\vT$ of the extracted electron is also of theoretical and experimental interest. It provides a kind of photographic image of the electron wave function inside the atom \cite{DEMKOV, KONDRA, BORDAS-theo}. 
The fringes of this distribution have been observed \cite{BORDAS-exp} with fields of the order of a few hundred V/cm for atomic energy levels $n\sim20$, just below or above the saddle point.
A quadrupolar asymmetry in $\vT$ has also been observed when the atom is excited with a linearly polarized light 
\cite{BORDAS-quadrupole}. 

The present work is devoted to the case where the initial electron state has a transverse orbital angular momentum perpendicular to $\mathbf{E}$.  
For this case a \emph{dipolar} asymmetry is predicted \cite{ARTRU-CZ}, 
with nonzero $\langle\mathbf{v}_{\rm T}\rangle$ in the direction of $\langle\mathbf{L}\rangle\times\mathbf{F}$, where $\mathbf{F}=-e\mathbf{E}$ is the external force. This effect, pictured in Fig.1, will be referred to as the ``\vLF'' asymmetry. 

\paragraph{Analogous effect in hadron physics.}

In high-energy hadron physics, the production of a quark-antiquark pair ($q\bar q$) in a QCD string (or flux tube) has a strong similarity with field ionization \cite{ARTRU-CZ}: the string tension (or the chromo-electric field) extracts the $q$ and the $\bar q$ from the vacuum via a tunneling effect. Assuming that the $q\bar q$ pair is initially in a $^3{\rm P}_0$ state  (corresponding to the vacuum quantum numbers \cite{QUARKISTES}), an effect analogous to the \vLF asymmetry should take place. 
This mechanism, pictured in Fig.2, was introduced \cite{LUND} to explain the polarization of hyperons produced in proton-proton collisions. It was later used \cite{ARTRU-CZ,ARTRU-CZ-Y} for modeling the Collins effect \cite{COLLINS}, which is the main tool of quark polarimetry. 
However, an alternative tunneling model of $q\bar q$ pair creation, based on the \emph{Schwinger mechanism}%
\footnote{This mechanism allows spontaneous $e^+e^-$ pair creation in a static field $E\sim mc^3/(\hbar e)$.
}, 
yields no \vLF asymmetry \cite{ARTRU-CZ}.  Thus the question of a \vLF asymmetry in quark pair creation remains open. 
It is important to check whether the analogous asymmetry exists in atomic physics.

\paragraph{Purpose and layout of the paper.}

A preliminary study \cite{RS-A-Constantine,XA-ERS-2013} showed that the \vLF asymmetry exists for the 2P state of hydrogen, 
but with an alternating time-dependence due to the \emph{linear} Stark effect. 
However, the field necessary to ionize the 2P state is not attainable in laboratory. We have therefore extended our study  to states of large $n$. Among these states we take those of maximal  transverse angular momentum.   
This paper is organized as follows:
Section 2 is a brief review of the \emph{Stark states}, \emph{i.e.} the stationary states of the hydrogen atom in a weak electric field, and their factorization in two harmonic oscillator wave functions. 
Section 3 presents the states of maximal $|\langle L_y\rangle|$, decompose them in the Stark basis and study the interrelated oscillations of $\langle \Lv_{\rm T}\rangle$ 
and $\langle \Av_{\rm T}\rangle$,  $\Av$ being the Runge-Lenz vector. 
Section 4 reviews the field ionization of a single Stark state, using the \emph{Gamow state} description. 
The tunneling rate and the asymptotic form of the Gamow wave function  are calculated at leading order in $F$;
the phase of the latter matters for the \vLF asymmetry.
In Section 5 are studied  the \vLF asymmetry for the ionization of states of maximal $|\langle L_y\rangle|$, and the conditions for having a sizeable asymmetry.
The conclusion is made in Section 6.

\section{Review of the Stark states} 

We consider a hydrogen atom in a static electric field $\mathbf{E}=-\mathbf{F}/e$ pointing in the $-\zu$ direction.
The Hamiltonian is $H=H_{0}-Fz$, where 
$H_{0}=\mathbf{p}^2/(2m_e)-\hat\alpha/r$ and $\hat\alpha= \alpha\hbar c$, ~$\alpha\simeq1/137$. We use the atomic units%
\footnote{In particular, the atomic units for time and force are $2.42\ 10^{-17}$ s and $5.14 \ 10^9$ eV/cm.
} (a.u.) in which $\hbar=m_e=\hat\alpha=1$. 
We will neglect the relativistic and radiative effects, in particular the spin-orbit coupling, the radiative widths and the Lamb shift. We assume $F\ll1$. Using the parabolic coordinates%
\footnote{Following the usual convention,  
$\xi$ and $\eta$ are respectively the "uphill" and "downhill" coordinates, but we take the $z$-axis in the "downhill" direction. 
} %
$\xi\! =\!r\!-\!z$, $\eta =r+z$, $\varphi =\arg \left( x+iy\right)$,  
the eigenstates of $H$ and $L_z$ are of the separable form \cite{LANDAU,FRIEDRICH}  
\begin{equation}  \label{sep} 
\Psi(\rv) = C\, \xi^{-1/2} \, \fxi(\xi) \ \eta^{-1/2} \, \feta(\eta) \,e^{im\varphi} \,.
\end{equation}
$C=2|\en|\sqrt\pi$  is a normalization coefficient, $\,\en$ is the energy. $\fxi$ and $\feta$ obey 
\begin{subequations} 
\begin{eqnarray}
\frac{\partial^2\fxi}{\partial \xi ^2}+\left[ \frac{\en}{2}+\frac{Z_\xi }{\xi }
-\frac{m^2-1}{4\xi ^2}-\frac{F\xi }{4}\right] \fxi(\xi) &=&0\,,
\label{eqsep-a}  
\\
\frac{\partial^2\feta}{\partial \eta^2}+\left[ \frac{\en}{2}
+\frac{Z_\eta }{\eta }-\frac{m^2-1}{4\eta^2}+\frac{F\eta}{4} \right] \feta(\eta) &=&0 \,,
\label{eqsep-b}  
\end{eqnarray}
\end{subequations}
with $Z_\eta+Z_\xi=1$. We set $\nu= (-2\en)^{-1/2}$,
$\sqrt{\xi/\nu}=\hat r$, $\sqrt{\eta/\nu}=\hat R$,
$\hat r \, e^{i\varphi} = {\hat x} + i \,{\hat y}$ and 
$\hat R \, e^{i\varphi} = {\hat X} + i {\hat Y}$. The functions 
\be \label{hatphi} 
\hat\fxi({\hat x}, {\hat y})  =  \xi^{-1/2}\fxi(\xi) \, e^{im\varphi}\,,\quad 
\hat\feta({\hat X}, {\hat Y})  =\eta^{-1/2}\feta(\eta) \,  e^{im\varphi}
\ee
are wave functions of two 2-dimensional \emph{anharmonic} oscillators \cite{GRAFFI}, obeying
\begin{subequations} 
\begin{eqnarray}
\left[2\exi+\Delta - {\hat r}^2 -\nu^3F\,{\hat r}^4\right] \hat\fxi({\hat x}, {\hat y})  &=&0\,,
\label{harm-a} 
\\
~ \left[2\eeta+\Delta - {\hat R}^2 +\nu^3F\, {\hat R}^4\right]  \hat\feta({\hat X}, {\hat Y}) &=&0\,.
\label{harm-b}  
\end{eqnarray}
\end{subequations}
They have the same angular momentum $m$ and their "energies"%
\footnote{
$\exi$ and $\eeta$ are twice $\beta_1$ and $\beta_2$ of Ref.\cite{YAMABE}.
},
$\exi=2\nu\,Z_\xi$ and $\eeta=2\nu\,Z_\eta$, are linked by $\exi+\eeta=2\nu$. 
Equation (\ref{sep}) rewrites
\begin{equation}  \label{resep} 
\Psi(\rv) = C\, \hat\fxi({\hat x}, {\hat y})  \, \hat\feta({\hat X}, {\hat Y}) \, e^{-im\varphi} \,.
\end{equation}
We will also use the mixed representation
\begin{equation}  \label{mix} 
\Psi(\rv) = C\, \hat\fxi({\hat x}, {\hat y})  \, \feta(\eta) / \sqrt\eta\,.
\end{equation}
\emph{Stark states} $|n_\xi,n_\eta,m\rangle$ are eigenstates of $H$ at lowest order in $F$, neglecting ionization.
$n_\xi$ and $n_\eta$ (usually denoted $n_1$ and $n_2$) are the numbers of nodes of $ \fxi(\xi)$ and $ \feta(\eta)$. The Stark wave functions are obtained putting for $\hat\fxi$ and $\hat\feta$ in (\ref{resep}) the 2-D \emph{harmonic} oscillator wave functions $\os$ \cite{COHEN-T}:
\be \label{fosc} 
\hat\fxi \Rightarrow \os_{\np,\nm}(\hat x,\hat y)=(\pi\, \np!\, \nm!)^{-1/2}  \, (a_+^\dagger)^{\np}
 \, (a_-^\dagger)^{\nm} \, e^{-(\hat x^2+\hat y^2)/2} ,
\ee
where the operator $a_\pm^\dagger = [{x-\partial_x\pm i (y-\partial_y)}]/2\,$ creates one quantum of clockwise ($-$) or anti-clockwise (+) excitation. $\,\np,\nm$ are the numbers of these quanta. 
Similarly, $\hat\feta \Rightarrow \os_{\NP,\NM}(\hat X,\hat Y)$. 
In the $F\to0$ limit, $\en\to-1/(2n^2)$ and    
\be \label{energ-parab} 
\begin{array}{l}
\nu\to n=n_\eta+n_\xi+|m|+1 \,,
\\
\exi\to \np+\nm+1 = 2n_\xi+|m|+1\,, 
\\
\eeta\to \NP+\NM+1 = 2n_\eta+|m|+1\,.
\end{array}
\ee
$C$ in Eqs.(\ref{sep},\ref{resep},\ref{mix}) is chosen such that $\langle\Psi|\Psi\rangle=1$ for Stark states.
For $n$=2 the Stark waves functions are
\be  \label{n=2wf} 
\begin{array}{rcl} 
 \Psi_{0,1,0} =& {\cal N} \, (\eta-2) \, e^{-(\xi+\eta)/4} &\equiv \Psi_1
\\
\Psi_{0,0,\pm1} =& {\cal N} \, \sqrt{\xi\eta} \, e^{-(\xi+\eta)/4} \, e^{\pm i\phi} &\equiv \Psi_{2\pm}
\\
 \Psi_{1,0,0} =& {\cal N} \, (\xi-2) \, e^{-(\xi+\eta)/4} &\equiv \Psi_3 \,,
\end{array}
\ee
with ${\cal N}=8^{-1}\pi^{-1/2}$. They correspond to $\exi=1,2,3$ respectively. We will use the simplified notation on the right. 

%
\begin{figure} [b!] 
  \centering
\includegraphics*[angle=90, bb= 90 80 500 780, width=0.8\textwidth]{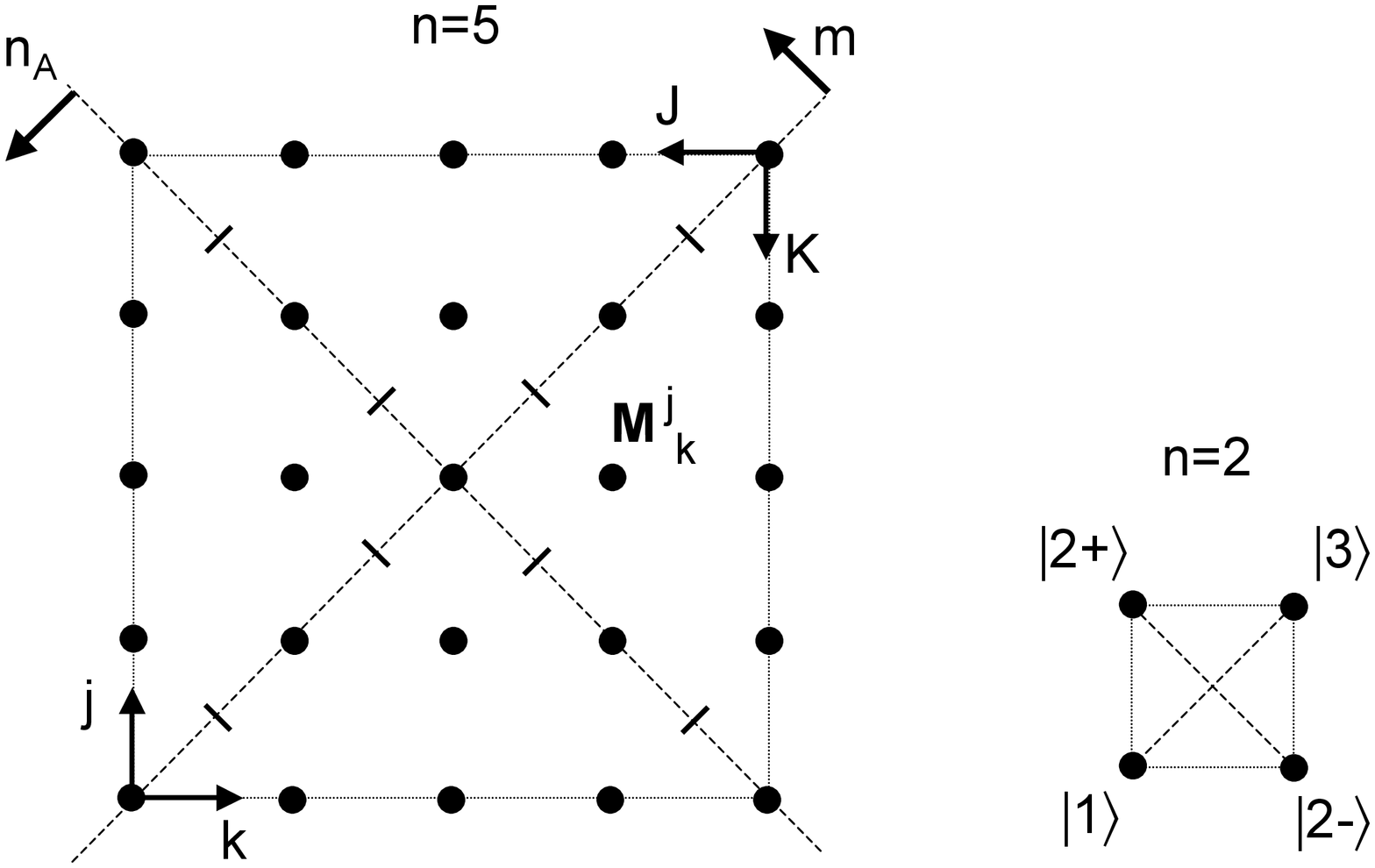} 
  \caption{\footnotesize 
The $\{j,k\}$ lattices of Stark states for $n$=5 and 2.  
$n$, $m$, $n_\xi$, $n_\eta$, $\np$, $\nm$, $\NP$, $\NM$ and $\nA$ are related by $\np-\nm=\NP-\NM=m$,~
$\np +\NM=\nm +\NP=n-1$,~
$n_\xi=\inf(\np,\nm)$,~
$\ n_\eta=\inf(\NP,\NM)$,~
$j+k+1=n-\nA$,~
$J+K+1=n+\nA$..
The $n$=2 states are those of Eq.(\ref{n=2wf}).
}
\end{figure}
%

Among the quantum numbers $n$, $m$, $n_\xi$, $n_\eta$, $\np$, $\nm$, $\NP$, $\NM$, only 3 are independent: we choose them to be $\{n, \np,\nm\}$. We also introduce $\nA\equiv n_\eta-n_\xi$. For fixed $n$ we represent a Stark state by a point $\Mv^{\np}_{\nm}$ in a square $n\times n$ lattice%
\footnote{
The quantities $\tilde m_{1}=j-(n-1)/2$ and $\tilde m_{2}=J-(n-1)/2$ 
 are the $z$ components of the pseudospins $j_1$ and $j_2$ of a $SO(3)\times SO(3)$ group \cite{BELLOMO}.
}
as shown in Fig.2. The relations between the above numbers are provided in the caption.
We will latter use the notation $|n,\Mv\rangle$ for $|n_\xi,n_\eta,m\rangle$.
To first order in $F$, the shifts of $\,\en$, $\lambda$ and $\mu$ are  
\be \label{deltaE} 
\begin{array}{rl}
\delta\en  = & - \nA\, \omega \ \ \text{with} \ \ \omega = 3Fn /2 \,,
\\
\delta \exi = & + Fn^3\, (3\exi^2-m^2+1)/4 \,, 
\\
\delta \eeta = & - Fn^3\, (3\eeta^2-m^2+1)/4 \,.
\end{array}
\ee

For large $n$, it will be useful to consider the Runge-Lenz vector  
\be
\Av =  \rv/r + (\Lv\times\pv-\pv\times\Lv)/2 \,, 
\ee
which is conserved for $F=0$, both in classical and quantum mechanics \cite{PAULI}
(other conventions exist for the sign and the normalization of $\Av$). 
A Stark states is an eigenstates of $A_z$ with the eigenvalue $\nA/n$.
$\,\Av$ and $\Lv$ are related through
\be  \label{ALips}
\Av\cdot\Lv = 0, \quad \Av^2 - 2\en \,(\Lv^2+\hbar^2) = 1 \,. 
\ee

\section{
Circular \boldmath $L_y$ eigenstates. 
Oscillations of $\langle L_y\rangle$ \unboldmath} 

We study the \vLF effect for states of extremal $L_{y}$, i.e., $L_{y}$=${\pm (n-1)}$.
They are represented semi-classically by circular orbits in the ($x,z)$ plane. 
Their waves functions are 
\begin{equation}   \label{Psi-n}  
\Psi _{Ly\pm } 
= [n^{n+1}\,l!\sqrt\pi]^{-1} \,  (z\pm ix)^{l} \, e^{-r/n}  \,, 
\end{equation}
with  $l=n-1$. For $n$=2, they are the following combinations of the Stark states (\ref{n=2wf}):%
\begin{equation}   \label{phiLy'} 
\Psi _{Ly\pm }=[\Psi _{1}-\Psi _{3} \pm i (\Psi _{2+} + \,\Psi _{2-})/2)] \,.
\end{equation}
The generalization to $n>$2 is
\be 
\label{decomp-n}  
\Psi _{Ly\pm} (\rv) 
=  \sum_{\Mv} c({\Mv}) \, \Psi_{n,\Mv} \,,  
\ee 
with (see Appendix A)  
\be \label{cM}  
c({\Mv})  
= (\pm i)^{\np+\nm} \, 2^{-l}  \left(C_l^{\np} \, C_l^{\nm} \right)^{1/2} .
\ee
$C_l^{k}\equiv l!/[k!(l-k)!]$ are the binomial coefficients. %
At large $n$, the largest $c({\Mv})$'s are reached for %
$\Mv$ near the center of the lattice. This is expected, since 
for a circular orbit in the $xz$ plane, we have classically $\L_z=A_z=0$. Quantum fluctuations give $\langle m^2\rangle = \langle \nA^2\rangle =l/2$. 

\begin{figure} [b!] 
  \centering
\includegraphics*[angle=+90, bb= 100 70 500 780, width=0.8\textwidth]{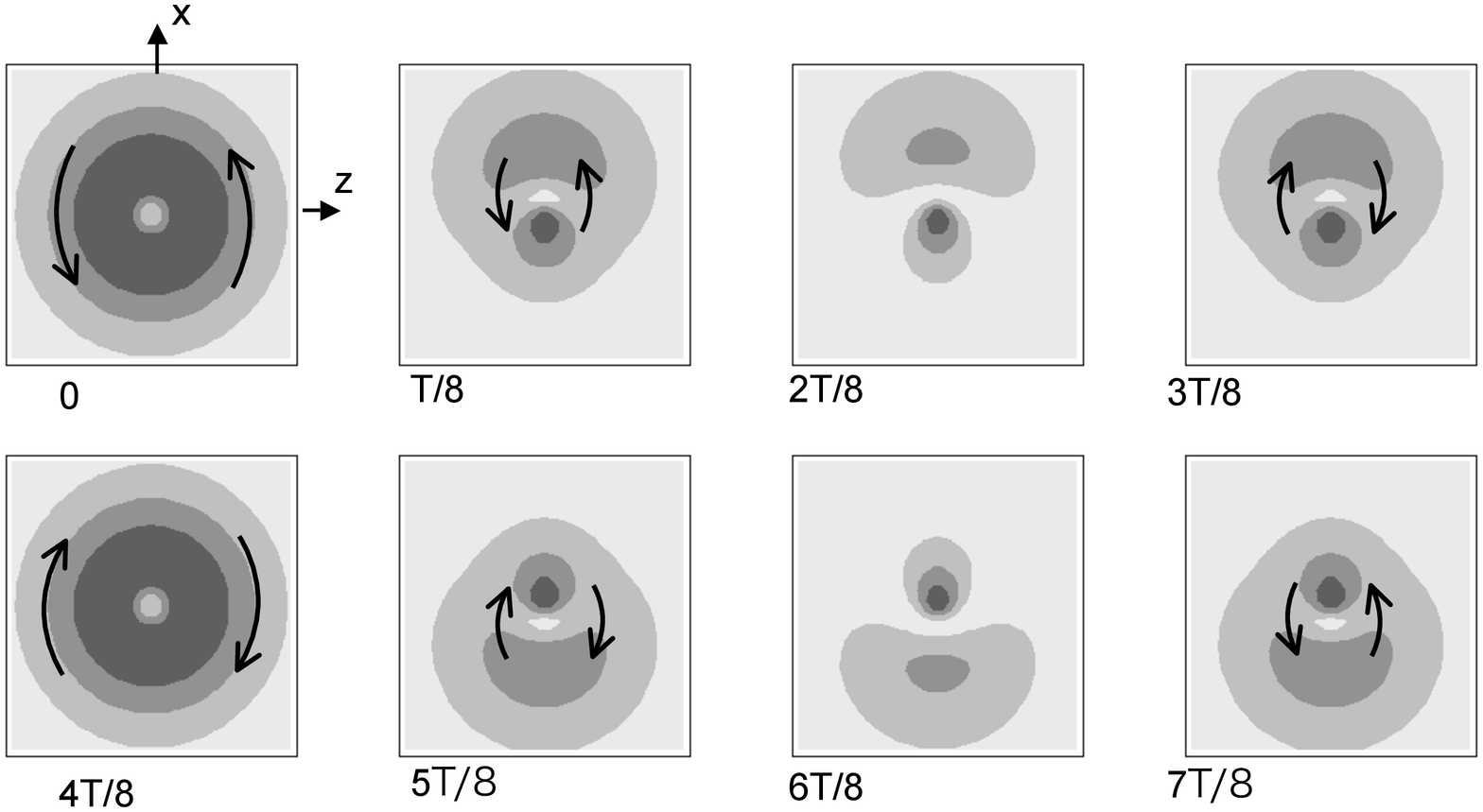} 
  \caption{ \footnotesize 
Evolution of the electron density $|\Psi(t,\rv)|^2$ of Eqs.(\ref{psiosc}-\ref{psiosc'}) during one oscillation period $T=2\pi/3F$ ($y=0$ slice). At $4t/T=0,1,2,3,4$, 
$\Psi(t)$ takes the values $\Psi _{Ly+}$, $i\Psi _{x+}$, $-\Psi _{Ly-}$, $-i\Psi _{x-}$ and $\Psi _{Ly+}$ again, where 
$\Psi _{x\pm}$ 
is the $A_x=\pm1/2$ eigenstate. 
$\Psi$ vanishes along the current vortex $z=0$,~ $2\pm x/\sin(3Ft)=r$ (whirling as indicated by arrows). 
For $t=T/4$ or $3T/4$, $\Psi$ vanishes on the surface $2\pm x=r$. 
The $x$- and $z$-windows are $[-7,+7]$. 
}
\end{figure}

Except for the $l=0$ states, $L_{y}$ eigenstates are not Stark states, therefore are mixed by the perturbing potential $-Fz$. Oscillations occur between them at a frequency equal to the Stark energy splitting $\omega=3Fn/2$. 
In the $n$=2 case, an atom initially in the state (\ref{phiLy'}) evolves according to%
\footnote{
$\,\Psi(t)$ is a short-hand notation for $\Psi(t,x,y,z)$ or $\Psi(t,\rv)$. When the argument $t$ is omitted, $\Psi$ designates 
the wave function at $t=0$:  ~$\Psi(\rv)\equiv\Psi(0,\rv)$.
}
\begin{eqnarray} \label{psiosc} 
\Psi(t) &=&e^{it/8} \left[ e^{+i\omega t}\, \Psi _{1} - e^{-i\omega t}\, \Psi _{3}
\,\pm\, i\,(\Psi _{2+} +\Psi _{2-}) \right]/2 
 \\
&=& e^{it/8}  \left[  \cos^2\frac{\omega t}{2} \, \Psi _{Ly\pm }
- \sin^2\frac{\omega t}{2} \,  \Psi _{Ly\mp }
+ \frac{i}{\sqrt2}\sin \left( \omega t\right) \Psi _{2S} \right] ,
 \label{psiosc'} 
\end{eqnarray}
where $\Psi _{2S} =  {\cal N}\sqrt2 (r-2) e^{-r/2}$ is the 2S wave function. Thus the atom oscillates between three $L_y$ eigenstates with a period $T= 2\pi/\omega$.
A complete oscillation of $\Psi(t,x,y,z)$ is analyzed in Fig.3.  
Eqs.(\ref{psiosc}-\ref{psiosc'}) ignore the decay of the states by tunneling ionization or by radiative transition.  

Oscillations of $\langle L_y\rangle$ for any $n$ are understood semi-classically with the help of the Runge-Lenz vector. 
Under the influence of the weak external force $\Fv$ a classical Kepler orbit varies slowly as depicted in Fig.4. $\Lv$ and $\Av$ are no longer conserved but evolve according to  
\begin{subequations} \label{L-A}  
\begin{eqnarray}
\langle d\Lv/dt\rangle  &=&  - [3/({4 |\en|})]\,  \Fv \times \langle\Av\rangle 
\,,
\label{L-A-a}  
\\
\langle d\Av/dt\rangle   &=& - (3/2)\,  \Fv \times \langle\Lv\rangle \,, 
\label{L-B-b}  
\end{eqnarray}
\end{subequations}
where the triangular brackets signify an average over time during one revolution.
Equation (\ref{L-A}) is also valid in the quantum mechanical case \cite{BELLOMO}.  It results from (\ref{L-A}) that $\langle L_y\rangle$ and $\langle A_x\rangle$ oscillate in quadrature, 
with the Stark frequency 
$\omega= 3nF/2$, drawing a circle in the $(L_y/n,A_x)$ plane in accordance with Eq.(\ref{ALips}). Such oscillations have been observed with Stark wave packets \cite{RAMAN}.

\begin{figure} [t] 
  \centering
\includegraphics*[bb= 10 610 680 800, width=140mm]{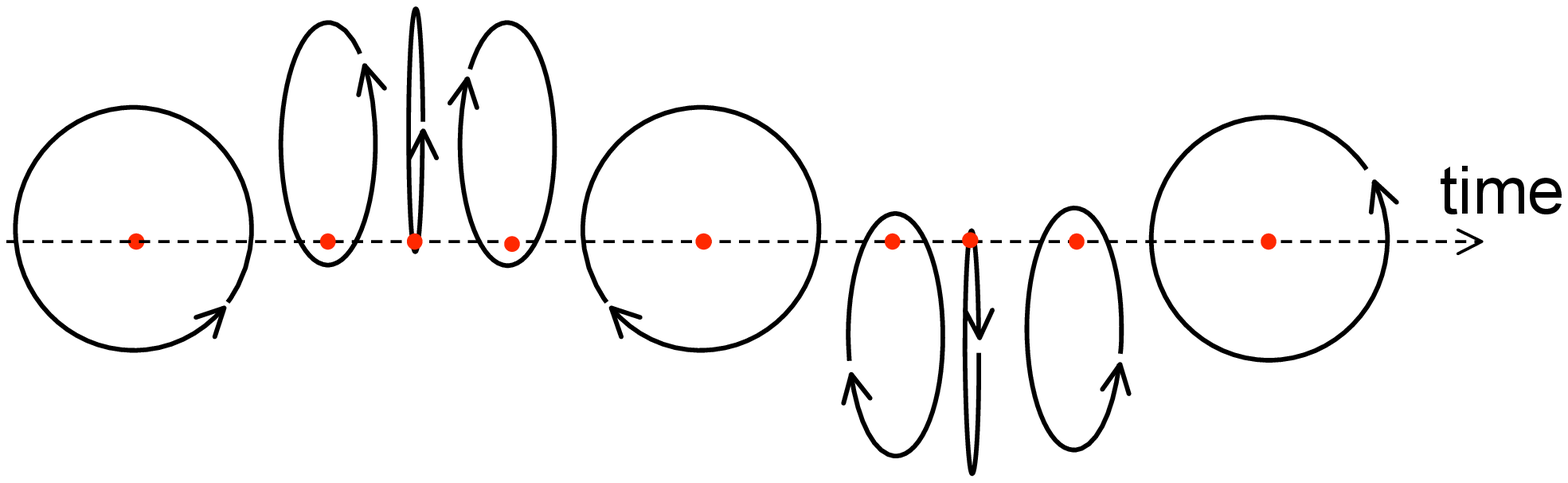} 
  \caption{\footnotesize 
Classical picture of the Stark oscillations of $L_y$ and $A_x$. The 8 first orbits correspond to the 8 panels of Fig.3.
}
\end{figure}

\paragraph{Experimental considerations.} 
As readily visible in Eq.(\ref{decomp-n}), circular $L_y$ eigenstates are coherent superposition of eigenstates of all $m$ values, $m$ ranging from -($n$-1) to ($n$-1). Therefore, their production is not trivial from an experimental point of view. Indeed, in a static electric field an optical excitation from a low excited state will populate only low $|m|$ values (typically $|m|$=0, 1 or 2). 
However there are methods \cite{DELANDE,BELLOMO} to transform adiabatically a ``blue" ($J=K=0$)  or ``red'' ($j=k=0$) Stark state into a circular $L_y$ eigenstate. 
In \cite{BELLOMO}, the electric field is rotated from the $+\xu$ or $-\xu$ direction to the $-\zu$ direction. An evolution like in Fig.4 between a needle-like and a circular trajectory takes then place. 

\section{Tunneling from Stark states}

For $F>0$, the eigenstates of $H$ remain bound in the $\xi$ variable. In the $\eta$ variable, they are continuous {scattering states}, containing both incoming and outgoing asymptotic waves. Stark states become \emph{Stark resonances}  which decay by tunneling ionization. 
As in most of the related works \cite{YAMABE,KOLOSOV}, we describe them as discrete states of complex energy $\en= \ER -i\gamma/2$, containing only the outgoing asymptotic wave: the \emph{Gamow states} \cite{GAMOW}. 

The ionization width, or decay rate, $\gamma$, of a Stark state has been calculated at lowest order in $F$ by Slavjanov \cite{SLAVJANOV}. For this calculation only the \emph{modulus} of asymptotic Gamow wave function $\Psi_{\rm G}$ is needed.  Here we are interested in the ionization of $L_y$ eigenstates, which are linear combination of Stark states, therefore we also need the \emph{relative phases} of the asymptotic $\Psi_{\rm G}$'s.  The semi-classical method giving both the widths and the relative phases is summarized below (details are given in Appendix):

The perturbation of $\fxi(\xi)$ by the external field is neglected. Tunneling bears on $\feta_{\rm G}(\eta)$. In the semi-classical approximation \cite{LANDAU},

\be  \label{Gamow-APP} 
\feta_{\rm G}(\eta) \simeq 
\feta_{\rm G}( \eta _{0})  \left[ {p_\eta(\eta _{0}) }/{p_\eta(\eta) }\right]^{1/2} 
e^{S(\eta_0,\eta)} \,,
\ee
\be   \label{K} 
S(\eta_0,\eta)= i  \int_{\eta_0}^{\eta}
p_\eta(s)\,ds \,,
\end{equation}
with $p_\eta^2(\eta) = \en/2 + {(1-m^2)/\eta^2} + Z_\eta/\eta + F\eta/4$.
$\,\eta_0$ is chosen deep inside the tunnel (classically forbidden region) and such that  $\feta_{\rm G}(\eta_0)$ can be approximated by the unperturbed Stark wave function $\feta_{\rm S}(\eta_0)$. The integration is done along a path in the upper complex half-plane, avoiding cuts of $p_\eta(\eta)$.
The result is independent of the precise choice of $\eta_0$.

\smallskip\ni\textbf{The width.}
Appendix B gives
\be \label{Slavjanov'} 
\gamma_{n,\Mv}= \left(\frac{4}{F\nu^3}\right)^{\eeta} \exp \left(-\frac{2}{3F\nu^3}\right)
\frac{\nu^{-3}}{\NP!\, \NM! } \,,
\ee
in accordance with Eq.(125) of \cite{YAMABE}. To lowest order in $F$ we replace $\eeta$ by $J+K+1=n+\nA$ 
and $\nu$ by $n$, except in the exponential where we use $\,\nu^{-3}=n^{-3}+9F\nA/2$. One gets the Slavjanov result \cite{SLAVJANOV}
\be \label{Slavjanov} 
\gamma_{n,\Mv}=\left(\frac{4}{Fn^3}\right)^{n+\nA} \exp \left\{-3\nA-\frac{2}{3Fn^3}\right\}
\frac{n^{-3}}{\NP!\, \NM!} 
\ee
reproduced in \cite{DAMBURG} and Eq.(126) of \cite{YAMABE} (noting that $\NP!\, \NM!\equiv n_\eta!\,(n-1-n_\xi)!$). 
It factorizes in the following way:
\be \label{Slavjanov-facto} 
\gamma_{n,\Mv} =  \gamma_{\rm min} \,  \left(\NP! \, f^{\NP}\right)^{-1} 
\, \left(\NM! \, f^{\NM}\right)^{-1} \,,
\ee
where $f\equiv Fn^3e^3/4$ and $\gamma_{\rm min}=n^{-3} \, f^{-1}  \, \exp\big\{ 3n-2/(3Fn^3)  \big\}$ is the width of the most stable sublevel ($\NP$=$\NM$=0). For the $n$=2 states  (\ref{n=2wf}),
\begin{subequations}  
\begin{eqnarray}   
\gamma_1&=& {2^{-6}F^{-3}} \ \exp \left[-{1/(12F)}-3\right] ,  
\\
\gamma_2  &=& {2^{-5}F^{-2}} \ \exp \left[ -{1/(12F)}\right] , 
\\
\gamma_3  &=& {2^{-4}F^{-1}} \ \exp \left[-{1/(12F)}+3\right] .
\end{eqnarray}   \label{gam} 
\end{subequations}

\smallskip\ni\textbf{The asymptotic Gamow wave function.}

The $\eta$-dependent phase is given by the second line of Eq.(\ref{Thetaa}). Replacing $\en$ by $-1(2n^2)+\delta\en-i\gamma/2$, $\,\eeta$ by $n+\nA$ in (\ref{Thetaa}), $\,\hat\fxi$ in (\ref{mix}) by $\os_{\np,\nm}$, and including the time dependence $e^{-i\en t}$ gives 
\be \label{Gas} 
\Psi_{\rm G}(t,\rv)  \sim  U (\te, \hat x, \hat y) \, V(t,\eta) \,,
\ee
\be \label{U} 
U( \te, \hat x, \hat y) = i^{n+\nA}  
\sqrt\gamma \, \exp\left\{- (\gamma/2 + i\delta\en) \te \right\}
\, \os_{\np,\nm}(\hat x,\hat y) \,,
\ee
\be \label{V} 
V(t,\eta) = \sqrt i \, (Fn^2\eta^3)^{-1/4} \,  
\exp\left\{it/(2n^2) + i (\eta-\bar\eta_{\rm ex})^{3/2} \sqrt F/3\right\} .
\ee
$\bar\eta_{\rm ex}= 1/(Fn^{2})$ is the the tunnel exit calculated for $\nA$=0, $|m|$=1, and $\te\equiv t-\sqrt{(\eta-\bar\eta_{\rm ex})/F}$ is the ``classical exit time". The exponential of (\ref{U}) takes into account the energy shift and the decay of the wave function, with the retardation effect due to the finite particle velocity. $V(t,\eta)$ is common to all states of given $n$.

$|\Psi_{\rm G}(t,\rv)|^2$ of (\ref{Gas}-\ref{V}) can be roughly interpreted as the density of a classical electron cloud falling freely in the uniform force field $\mathbf{F}$. An electron of this cloud leaves the tunnel at time $\te$ and follows approximately a parabolic motion $z = \bar\eta_{\rm ex}/2 + (t-\te)^2F/2=(x/\hat x)^2/(2n) = (y/\hat y)^2/(2n)$, at fixed $\hat x$ and $\hat y$. Its transverse velocity $\vv_\perp \simeq (\hat x, \hat y)\,  \sqrt{nF}$ gives access to $|\hat\fxi(\hat x,\hat y)|^2$ in the imaging method of \cite{DEMKOV, KONDRA, BORDAS-theo, BORDAS-exp,BORDAS-quadrupole}.

Our approximations rest on $F\ll F\crt$, where we define the \emph{critical field} $F\crt$ to be such that the unperturbed ``energy" in (\ref{harm-b}) equals the potential barrier: 
\be  \label{critic=} 
F\crt = [8 n^3(n+\nA)]^{-1} \,.   
\ee

\section{Decay of a \boldmath$L_y$ \unboldmath eigenstate and the \vLF asymmetry} 

We consider an atom occupying the $\Psi_{Ly\pm}$ state (\ref{Psi-n}) at $t\le0$ and placed in the external field $\Ev=F\,\zu$ at $t\ge\Delta t$. During the transitory period $[0,\Delta t]$ the field grows smoothly from 0 to $F$. We suppose $1/n^3\ll\Delta t\ll1/\omega$, so that non-adiabatic transitions to states of different $n$ can be neglected and the oscillations described in Section 3 have scarcely started at $t=\Delta t$. 
If we neglect tunelling ionization, the atom wave function evolves for $t\ge\Delta t$ according to the generalization of Eq.(\ref{decomp-n}),
\be 
\label{decomp-n-bis}  
\Psi(t,\rv) 
=  \sum_{\Mv} c({\Mv}) \, \Psi_{n,\Mv}(t,\rv) \,,  
\ee 
with
\be 
\label{decomp-n-ter}  
 \Psi_{n,\Mv}(t,\rv) 
=  \Psi_{n,\Mv}(\rv) \, e^{[1/(2n^2)+\nA\omega]\, it}  \,.  
 \ee 
and the oscillations of $L_y$ and $A_x$ given by (\ref{L-A}) take place. 
We now take tunelling ionization into account by replacing $\Psi_{n,\Mv}(t,\rv)$ in (\ref{decomp-n-bis}) by the corresponding Gamow wave function. At large $\eta$ and for $\te>0$ (see the note below) we use the asymptotic form (\ref{Gas}-\ref{V}) and obtain
\begin{subequations} 
\begin{eqnarray} \label{GaLy} 
\Psi(t,\rv)  &\sim& V(t,\eta) \, 
\profil (\te,\hat x, \hat y ) \,,
\\
\profil( \te, \hat x, \hat y)  &=& \sum_\Mv c({\Mv})  
\, U_{n,\Mv} (\te, \hat x, \hat y).  
\label{GaLy'} 
\end{eqnarray}
\end{subequations} 
$|\profil( \te, \hat x, \hat y)|^2$ may be measured with the imaging method of \cite{DEMKOV, KONDRA, BORDAS-theo, BORDAS-exp,BORDAS-quadrupole}.

\smallskip
\ni
\textbf{Note:} 
Once the external field is switched on, the transition from a Stark wave function $\Psi_{\rm S}(t,\rv)$ to a Gamow wave function $\Psi_{\rm G}(t,\rv)$ is not instantaneous. At large $\eta$, due to the finite velocity of the electron, the Gamow wave function sets up only when $\te$ becomes positive, i.e., 
 $t >\sqrt{(\eta-\bar\eta_{\rm ex})/F}$. 

\subsection{The \vLF asymmetry for n=2} 

Let us first study the simplest case $n$=2. It contains the basic features which remain at higher $n$. 
From (\ref{GaLy'},\ref{U},\ref{cM}), 
\begin{equation}  \label{profile} 
\begin{aligned} 
\profil(\te, \hat x, \hat y)  &= (-i/2) \, \big\{
\sqrt{\gamma_1} \,  e^{(i\omega-\gamma_1/2)\te} \, \os_{00}  
\\
& + \sqrt{\gamma_3} \, e^{(-i\omega-\gamma_3/2)\te} \, \os_{11} 
\pm \sqrt{\gamma_2} \, e^{-\gamma_{2} \te/2} \, ( \os_{10} +  \os_{01} )
\big\} .
\end{aligned} 
\end{equation}
Applying (\ref{fosc}) and squaring, 
\begin{equation}  \label{QXY} 
\begin{aligned} 
|\profil(\te, \hat x, \hat y)|^2 & = (4\pi^{-1}) \, e^{-\hat x^2 - \hat y^2} \big\{
 \gamma_1
 \, e^{-\gamma_1\te}  
 +  4\gamma_2 \, e^{-\gamma_2\te} \, \hat x^2
\\
&+  \gamma_3 \, e^{-\gamma_3\te} \, ({\hat x}^2+{\hat y}^2-1)^2 
\\
& \pm  4\sqrt{\gamma_1\gamma_2} \, e^{-\bar\gamma_{12} \te} \, 
\hat x \, \cos(\omega \te)  
\\
& \pm 4 \sqrt{\gamma_2\gamma_3} \, e^{-\bar\gamma_{23} \te}  
\, \hat x \, ({\hat x}^2+{\hat y}^2-1) \cos(\omega \te)  \\
 & + 2\sqrt{\gamma_1\gamma_3} \, e^{-\bar\gamma_{13} \te}  \, \cos(2\omega \te)
 \, ({\hat x}^2+{\hat y}^2-1) 
 \big\}.
\end{aligned} 
\end{equation}
$\gamma_i$ is given by (\ref{gam}) and $\bar\gamma_{ij}\equiv(\gamma_i+\gamma_j)/2$.    
The 1,2 and 2,3 interference terms are odd in $\hat x$. They give a \vLF asymmetry, recalling that $ v_x \simeq \hat x \,  \sqrt{nF}$. 
As a measure of the asymmetry, we define
\be   \label{V?} 
a(\te) \equiv \langle v_x\rangle / \Delta v_x   =
I^{(1)} (\te) \, \left[ {I^{(0)} (\te)\,I^{(2)} (\te)} \right]^{-1/2}  \,,
\ee
\be   \label{M} 
I^{(q)} (\te) \equiv \langle \profil(\te,\hat x, \hat y)\,| \hat x^q| \,\profil(\te,\hat x, \hat y) \rangle\,.
\ee
From Eq.(\ref{QXY}), 
\begin{subequations}  \label{Mi}
\begin{eqnarray} \label{M0}
& 4 I^{(0)}  = 
\gamma_1 \, e^{-\gamma_1 \te} + 2 \gamma_2 \, e^{-\gamma_2 \te} + \gamma_3 \, e^{-\gamma_3 \te} \,, 
\\
& 2 I^{(1)} = 
\pm  \big[
\sqrt{\gamma_1\gamma_2} \, e^{-\bar\gamma_{12} \te} +
\sqrt{\gamma_2\gamma_3} \, e^{-\bar\gamma_{23} \te}  
\big]  \, \cos(\omega \te) \,,
\label{M1}  
\\
& 8 I^{(2)} = 
\gamma_1 \, e^{-\gamma_1 \te} + 6 \gamma_2 \, e^{-\gamma_2 \te} +2\sqrt{\gamma_1\gamma_3} \, e^{-\bar\gamma_{13} \te}  \, \cos(2\omega \te) 
+ 3 \gamma_3 \, e^{-\gamma_3 \te} .
\label{M2}  
\end{eqnarray}
\end{subequations}
An alternative measure of the asymmetry is 
\be\label{aalt}
a_{\rm alt}(\te) = I^{(0+)} / I^{(0)} \,,
\ee
where $I^{(0+)}$ is the integral of $ |\profil(\te, \hat x, \hat y)|^2$ restricted to the half-plane $\hat x>0$.
%
\begin{figure}
\centering
\includegraphics*[bb= 0 600 550 850, width=80mm]{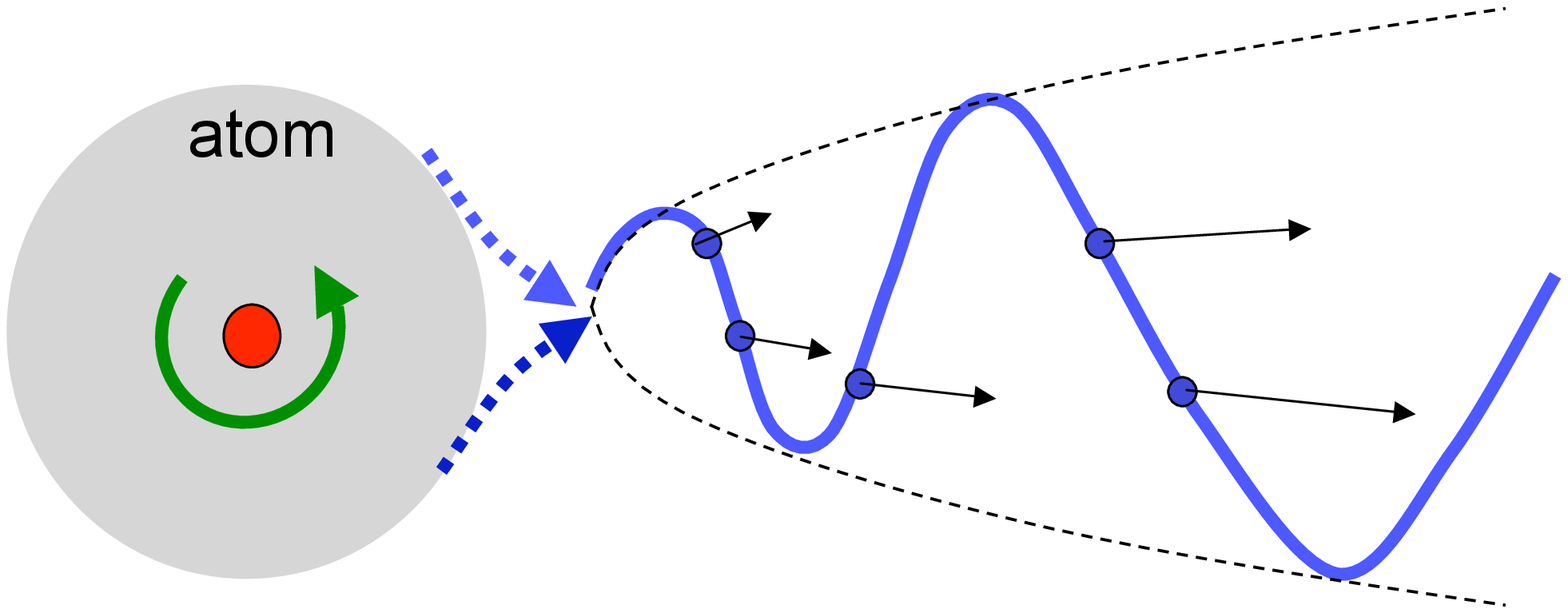}
\caption{\footnotesize 
Motion of the probability density $|\Psi(t\rv)|^2$ of the outgoing electron. 
 The curve represents $\langle x(t,z)\rangle \simeq 2 \langle \hat x(t) \rangle \sqrt{z}$ versus $z$ at fixed $t$. As $t$ increases the undulations move to the right. It looks like a crawling snake. 
}
\end{figure}

\paragraph{The time-dependent asymmetry.}   
The initial asymmetry,
\be 
\bigg\{
\begin{array}{rcl}
 \label{a0} 
a(0) &=& \pm \, 8^{1/2} \left(f^{-1} +8+ 3f  \right)^{-1/2} 
\,,
\\
a_{\rm alt}(0) &=& \pm \, (\pi f)^{1/2} \, (1+f/2) / (1+2f)^2 \,,
\end{array}
\ee 
is in the direction indicated in Fig.1. 
At $\te>0$, $\, a(\te)$ oscillates at the Stark frequency $\omega=3F$ and in phase%
\footnote{It looks as if the group velocity is infinite in the tunnel.}  
with $\langle L_y\rangle$. The outgoing electron current is pictured in Fig.5.

The outgoing flux is proportional to $I^{0}$. As pictured in Fig.6, it does not follow a simple exponential decay and $a(\te)$ does not oscillate with constant amplitude.
We denote by $a^{\rm sup}(\te)$ the {upper envelope} of $a(\te)$, obtained by replacing $\cos(\omega\te)$ and $\cos(2\omega\te)$ by 1 in (\ref{Mi}).
 Considering first the case $F\ll F\crt$, hence $\gamma_1 \gg  \gamma_2 \gg  \gamma_3$, we distinguish three ``eras'', depending on which Stark state gives the main contribution to (\ref{Mi}a,c). The eras change at times $\te_{12}$ and $\te_{23}$ given by  $ \gamma_i \,  e^{\gamma_i \,\te_{ij}} = \gamma_{j} \,  e^{\gamma_{j} \,\te_{ij}}$ or
\be
\te_{ij} = (\gamma_i - \gamma_{j})^{-1} \, \ln(\gamma_i/\gamma_{j}) \,.
\ee
\smallskip  
\ni 1) \emph{the $\Psi_{1}$ era}, $[0,\te_{12}]$: 
 $a^{\rm sup}(\te)$ starts from the small value $a(0)\sim\sqrt{8f}$, then increases like $e^{(\gamma_1-\gamma_2)\te/2}$ up to $\sim\sqrt{8/21}$
at time $\te_{12}$ where $\Psi_{1}$ and $\Psi_{2}$ interfere with maximum efficiency. 

\smallskip
\ni 2) \emph{the $\Psi_{2}$ era}, $[\te_{12},\te_{23}]$: %
the $\Psi_{1}$ component has almost decayed. 
$I^{(0)}$ and $I^{(2)}$ are dominated by the second terms of (\ref{Mi}a,c), while $I^{(1)}$ is dominated  successively by the 12 and 23 terms. $a^{\rm sup}(\te)$ decreases, then increases again up to $\sim\sqrt{8/27}$. 

\smallskip
\ni 3) \emph{the $\Psi_{3}$ era}, $[\te_{23},\infty]$: the $\Psi_{2}$ component also has almost decayed.
(\ref{Mi}a,b,c) are dominated by their last terms. 
$a(\te)$ decreases like $e^{(\gamma_3-\gamma_2)\te/2}$. 
%
%
\begin{figure}   [t]
\centering
\includegraphics*[angle=90, bb= 30 150 530 680, width=80mm]{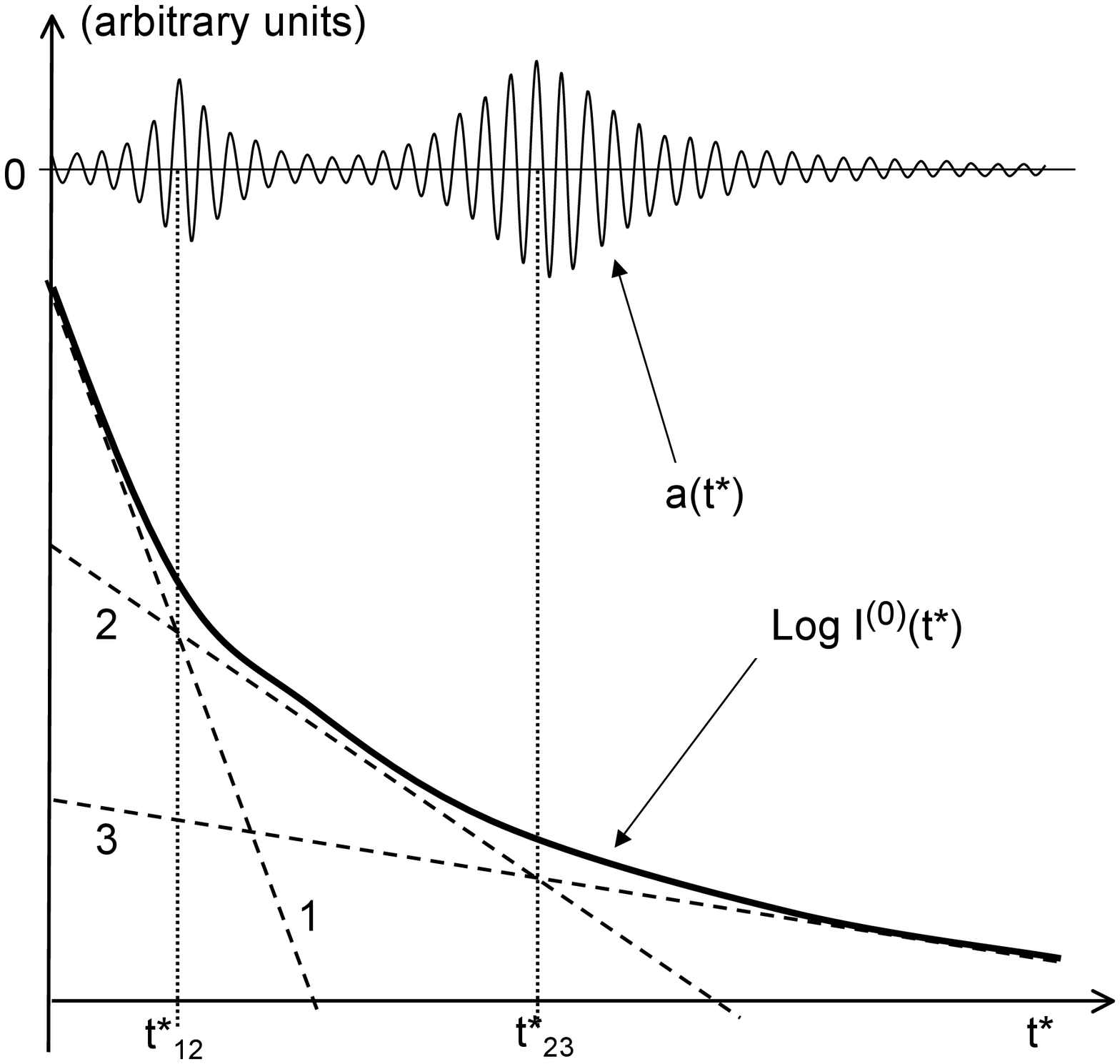} 
\caption{\footnotesize
Time dependence of $\log I^{(0)}$, $\,I^{(0)}$ given by Eq.(\ref{M0}) being the outgoing electron flux, and of the asymmetry parameter $a$ given by Eq.(\ref{V?}), in the regime $F\ll F\crt$. The dashed lines labeled $i$=1,2,3 are the curves $\log(\gamma_i \, e^{-\gamma_i \te})$ corresponding to individual Stark states. The vertical lines delimit the successive ``eras''.  
}
\end{figure}  

\paragraph{The time-averaged asymmetry.}  

Except for very small values of $F/F\crt$, $\,a(\te)$ oscillates too fast to be measured with an ordinary detector. 
For instance $F>0.002$ gives $T=2\pi/3F\lesssim 10^3$ a.u. $\sim 10^{-14}\,$s. Therefore one can only measure a time-averaged asymmetry
\be \label{<a>}
\langle a\rangle =   I^{(1)}_{\rm int} \left[ I^{(0)}_{\rm int}\, I^{(2)}_{\rm int} \right]^{-1/2},
\ee 
where  $I^{(q)}_{\rm int}$ is the $\te$-integrated value of $I^{(q)}$. We have $I^{(0)}_{\rm int}=1$,
\be \label{Mint}
\begin{aligned}
I^{(1)}_{\rm int} &=  \frac{\pm \sqrt{2f} }{1+f}\, \left(
\frac{\bar\gamma_{12}^2}{\bar\gamma_{12}^2+\omega_{12}^2}
+ \frac{\bar\gamma_{23}^2}{\bar\gamma_{23}^2+\omega_{23}^2}
\right),
\\
I^{(2)}_{\rm int} &=  \left( \frac{5}{2}+ \frac{f}{1+f^2}
\, \frac{\bar\gamma_{13}^2}{\bar\gamma_{13}^2+\omega_{13}^2} \right) .
\end{aligned}
\ee
If $\gamma_1\ll\omega$, $\, \langle a\rangle$ is suppressed due to the oscillations of $I^{(1)}$. To prevent this suppression, one must  take $F$ such that the Stark oscillations are quenched by the decay of the $\Psi_1$ state, that is $\gamma_1 \gtrsim \omega=3F$. %

\begin{table}
\caption{
\footnotesize{Numerical results for the widths (Eq.\ref{gam}), the initial asymmetry (Eq.\ref{a0}), and the time-averaged asymmetry (Eqs.\ref{<a>}-\ref{Mint}), to lowest order in $F$. In brackets under our figures are quoted exact numerical values for $\gamma$, from
$^a$Ref.\cite{DAMBURG}(1978), $^b$Ref.\cite{DAMBURG}(1976),  $^c$Ref.\cite{KOLOSOV}.}
}  
\begin{center}
\begin{tabular}{llllllll}
\hline\hline
~F & $~\gamma_1$ & $~\gamma_2$ & $~\gamma_3$ & $a(0)$ 
& $a_{\rm alt}(0)$& $~\langle a\rangle$ 
\\
\hline \\
0.006 & 0.0033 & 0.0008 & 0.0002 & 0.79 & 0.44 & 0.0049
\\[-0.5ex]
 & & & (0.000061)$^a$ & &
\\[1ex]
0.0065 & 0.0077 & 0.0020 & 0.00052 &  &  &  
\\[1ex]
0.008 & 0.045 & 0.015 & 0.0047 & 0.81 & 0.43 & 0.28 
\\[-0.5ex]
 & (0.0042)$^b$ & (0.0020)$^b$ & (0.00085)$^b$ & &
\\[1ex]
0.010 & 0.19 & 0.075 & 0.030 &  0.83 & 0.41 & 0.66
\\[-0.5ex]
 & (0.011)$^c$ & (0.0063)$^c$ & (0.0033)$^c$ & & \\[1.5ex]
\hline
\end{tabular}
\end{center}
\end{table}
\paragraph{Numerical results for \boldmath$n$\unboldmath=2 and discussion.}

Table 1 shows the numerical results of the above formulae for four values of the field. The initial values of $a(\te)$ and $a_{\rm alt}(\te)$ are also given. For the two largest values of $F$, we have $\gamma_1>\omega=3F$, Stark oscillations are quenched and $\langle a \rangle$ is sizeable. The widths given by Eq.(\ref{gam}) are much larger than those at all orders in $F$ given in literature. This is because the chosen $F$ are not small compared to $F\crt$ (=0.0078 for $\Psi_2$). 
Nevertheless we believe that the main results of this subsection are qualitatively true. Indeed, the discrepancy between the widths calculated at of all-order and lowest-order in $F$ is strongly reduced if we re-calculate the latter after a small reduction of $F$. This is due to the very steep slope of the curve $\log\gamma$ versus $F$. 
Compare, for instance, the all-order results for $F=0.008$ with those of Eq.(\ref{gam}) for $F=0.0065$.

\subsection{\vLF asymmetry at large \boldmath $n$ \unboldmath} 

The $n$=2 case was instructive but the field required for ionization is not attainable in laboratory. We therefore explore the large $n$ case, where the critical field $\sim1/(8n^4)$ is considerably reduced and experiments become feasible. The $n$=2 results generalize as follows: 

The $\Mv^{\np}_{\nm}$ term of (\ref{GaLy'}) contains the factor $e^{im\phi}$, with  $m=\np-\nm$. Because $c(\Mv^{\np}_{\nm}) = c(\Mv^{\nm}_{\np}) $ the sum of the $\Mv^{\np}_{\nm}$ and $\Mv^{\nm}_{\np}$ terms contains the factor $\cos m\phi$ and is of parity $(-1)^m$ in $\hat x$. The interferences betwen even- and odd-$m$  terms give the \vLF asymmetry.  Let us measure the latter using $a(\te)$ of  (\ref{V?}-\ref{M}). 
Gathering (\ref{U},\ref{GaLy'} and \ref{M}), we have
\be  \label{Iq-app} 
\begin{aligned}   
I^{(q)}(\te) =& \sum_{\Mv,\Mv'} (\pm1)^{\nA-n'_{\rm A}} \,
|c({\Mv}) \, c(\Mv')|  \, (\gamma_{\Mv} \gamma_{\Mv'})^{1/2}
\, \langle \os_{j',k'}| \hat x^q | \os_{j,k}\rangle 
\\
&\exp\{-(\gamma_{\Mv} +\gamma_{\Mv'})\te/2\} \,
\cos \big[ (n'_{\rm A}-\nA) \omega \te  \big] \,,
\end{aligned}  
\ee
where $\gamma_{\Mv}$ stands for $\gamma_{n,\Mv}$. 
For $I^{(0)}$ the orthogonality of the $\os_{j,k}$'s
imposes $\Mv=\Mv'$. 
For $I^{(1)}$ we use (\ref{fosc}) and $2\hat x = a_+ + a_- + a_+^\dagger + a_-^\dagger$, from which $|j'-j|+|k'-k|=1$. For $I^{(2)}$, $|j'-j|+|k'-k|=0$ or 2. 
More generally, the summation in (\ref{Iq-app}) is restricted to $\Mv\Mv'$ pairs connected by a walk of $q$ horizontal or vertical steps n Fig.2.
The useful matrix elements are 
\begin{equation}  \label{elematr} 
\begin{aligned} 
& \langle j-1,k| \hat x^1 | j,k\rangle = \sqrt j/2 \,,
\\
& \langle j,k| \hat x^2 | j,k\rangle = (j+k+1)/2 \,,
\\
& \langle j-1,k-1| \hat x^2 | j,k\rangle = \sqrt{jk}/2 \,, 
\\
& \langle j-2,k| \hat x^2 | j,k\rangle = \sqrt{j(j-1)}/4 \,,
\end{aligned} 
\end{equation}
and symmetric expressions with $j\leftrightarrow k$. 
Two $(\Mv,\Mv')$ pairs which deduce from each other by a $\Mv\leftrightarrow \Mv'$ or $j\leftrightarrow k$ permutation give identical contributions to $I^{(q)}$.

\smallskip\ni\textbf{Time-dependent asymmetry.}
For large enough $n$, hence a weak ($\sim n^{-4}$) electric field and a small ($\sim  n^{-3}$) Stark frequency, a time-resolved experiment could be possible. For instance in \cite{NOORDAM}, the escaping times of the electrons are recorded with a picosecond resolution, for $n\sim20$. Inserting a half-plane screen, one could measure $a_{\rm alt}(\te)$ defined by Eq.(\ref{aalt}).

As in the $n$=2 case, $a(0)\sim f^{1/2}$. All the contributions to $I^{(1)}(0)$ have the same sign, giving an asymmetry oriented as in Fig.1. 
At large $n$ and for small field ($F\lesssim F\crt^{(\nA=0)} = n^{-4}/8$), 
we have $f \lesssim n^{-1}$ 
therefore the main contribution to $ I^{(q)}(0)$ comes from the decay of the most unstable states (largest $\NP$ and $\NM$) although they have small coefficients in the expansion (\ref{decomp-n}). These states decay first, then the outgoing flux is successively fed by more and more stable states, generalizing the $n$=2 scenario illustrated by Fig.6.
Peaks of the oscillation amplitude of $a(\te)$ occur at times
\be
\te_{\Mv\Mv'} = (\gamma_{\Mv} - \gamma_{\Mv'})^{-1} \, \ln(\gamma_{\Mv}/\gamma_{\Mv'}) \,,
\ee
for $\Mv$ and $\Mv'$ separated by one step in Fig.2. 

\smallskip\ni\textbf{Time-averaged asymmetry.}
In a more simple experiment one measures $\langle a\rangle$, defined by (\ref{<a>}), or a time-average of $a_{\rm alt}$ defined similarly. $\langle a\rangle$ is obtained replacing the second line of (\ref{Iq-app}) by 
\be
\frac{(\gamma_{\Mv}+\gamma_{\Mv'})/2}
{ (\gamma_{\Mv}+\gamma_{\Mv'})^2/4+(\nA-\nA')^2\,\omega^2}
\,.
\ee
To contribute significantly to $\langle a\rangle$, a $\Mv\Mv'$ pair must fulfill $\sup(\gamma_{\Mv},\gamma_{\Mv'}) \gtrsim \omega$. 
It must also be near the center of the lattice of Fig.2 ($\nA$ and $m\ll n$) to insure a large $|c({\Mv})c({\Mv'})|$. In this region we can apply the Stirling formula in (\ref{Slavjanov}). It gives 
\be \label{gammaStirling} 
\gamma_{\Mv}\simeq\frac{1}{2\pi n^3} \left(\frac{8e}{
Fn^3(n+\nA)
} \right)^{n+\nA}
\exp \left\{-3\nA-\frac{2}{3Fn^3} - \frac{m^2}{2(n+\nA)} \right\} \,.
\ee
To leading orders in $n$ and $F$, we have
\be \label{gam/om} 
\ln(\gamma_{\Mv}/\omega) 
\simeq n\,[\ln(64F\crt/F)+1-(16/3)F\crt/F]
\,,
\ee
with $F\crt\simeq 1/(8n^4)$.
We conclude that $\langle a\rangle$ is large only for $F\sim F\crt$ (we exclude the case $F> F\crt$).
In this case, the widths of the participating Starks states are comparable. We have indeed 
\be
\gamma_{\Mv'}/\gamma_{\Mv} \simeq 
[(e^3/64) \, F/  F\crt]^{\nA-\nA'} \,. 
\ee
Therefore we do not expect peaks of the oscillation as in the small $F$ case.

\paragraph{Comparison with exact numerical results at large \boldmath$n$\unboldmath.} 

We choose $F\sim F\crt\sim1/(8n^4)$, in order to get a large enough $\langle a\rangle$. 
As for $n$=2, the Slavjanov formula (\ref{Slavjanov}) greatly overestimate the widths.  
Let us consider for instance the case $n$=10 and $F$=10$^{-5}$.  For $m$=1 and $n_\xi$=4, Table 5 of Ref.\cite{LUC} gives $\gamma$=$3.31\,10^{-12}$ whereas Eq.(\ref{Slavjanov}) gives  $\gamma$=$4.06\,10^{-10}$ (here $J$=5, $K$=4 and $\nA$=0). This is because $F$ is not far from the critical field $F\crt=1.25\ 10^{-5}$.  
Again, the correct result can be recovered with a small reduction of $F$ ($F$=0.923~10$^{-5}$) in (\ref{Slavjanov}). 
Besides, the ratio $\gamma_{\Mv'}/\gamma_{\Mv}$ for neighbouring $\Mv$ and $\Mv'$
is reasonably well described by Eq.(\ref{Slavjanov-facto}). 
This is best seen using the quantity $\hat\gamma \equiv J!\,K!\,\gamma$.
According to Eq.(\ref{Slavjanov-facto}) the ratio between the $\hat\gamma$'s for two successive values of $n_\xi$ and $m=1$ is equal to $f^2\simeq1/400$ for $F$=10$^{-5}$. 
According to Table 5 of Ref.\cite{LUC}, this ratio ranges from 1/379 to 1/324.
Therefore, as in the $n$=2 case, we believe that our results are qualitatively valid.

A last point: the \vLF asymmetry greatly depends on the relative phases between asymptotic Gamow wave functions for neighbouring $\Mv$'s. The $\Mv$-dependent phase comes from the term $i\pi\eeta/2$ in (\ref{KetaT}) and we made the approximation $\eeta\simeq n+\nA$.   
We assume that the exact phase difference for neighbouring $\Mv$ and $\Mv'$ is not too different from $(\nA-\nA')\pi/2$.

\section{Conclusion}

We have theoretically established the \vLF asymmetry in strong-field tunneling ionization of a hydrogen atom with transverse orbital angular momentum. On the average, 
the extracted electron has a transverse velocity $\langle\vv_{\rm T}\rangle$ in the same direction as just before it entered the tunnel. 
For fields smaller than the typical critical field $F\crt\sim 1/(8n^4)$ (in a.u.), the linear Stark effect produces oscillations of $\langle \Lv_{\rm T} \rangle$, therefore of $\langle \vv_{\rm T} \rangle$, making the time-averaged $\langle a\rangle$ asymmetry very small. A time-resolved experiment is necessary.
 For $F\sim F\crt$, although our formulae become inaccurate, the prediction of a sizeable time-averaged asymmetry should be qualitatively correct.  

The \vLF asymmetry should exist as well in the strong-field ionization of other atoms. Its underlying mechanism may also be at work in the analogous effects (Collins effect and hyperon polarization) of hadron physics. A related effect may occur in the capture of an atomic electron by a ion: crossing the potential barrier, the electron should keep the orientation of its transverse momentum relative to the nucleus-nucleus axis. Then, its angular momentum in the new atom should be opposite to that in the initial atom.  

\subsection*{Acknowledgment}

We thank J.-M. Richard and Marjorie Shapiro for helping to improve the manuscript and Ch. Bordas for interesting informations and suggestions concerning the experimental aspects.

\section{Appendix A. Derivation of  Eq.(\ref{cM})} 

The coefficients $c(\Mv)$ 
can be obtained by fitting the asymptotic behaviors of  (\ref{Psi-n}) and (\ref{decomp-n}),
using the variables $u=\hat x+i\hat y=e^{i\phi}\sqrt{\xi/n}$, $\hat R=\sqrt{\eta/n}$. For (\ref{Psi-n}) we write
$z+ix= (\eta-\xi\pm i \sqrt{\eta\xi} \cos\phi)/2 = (n/2) (\hat R\pm i u) (\hat R\pm i\bar u)$, whence
\be  \label{binom} 
\Psi _{Ly\pm} (\rv) = [n^{2} l!\sqrt\pi]^{-1} \,2^{-l} 
\, \sum_{j=0}^l  C_{l}^j (\pm i  u)^{j} \,  \hat R^{l-j} \,
\, \sum_{k=0}^l   C_{l}^k \, (\pm i\bar u)^{k} \, \hat R^{l-k} \,,
\ee
with $l=n-1$. For $\Psi_{n,\Mv}$ in (\ref{decomp-n}),
we use (\ref{resep}), the asymptotic form of (\ref{fosc}),
\begin{eqnarray}   \label{fosca} 
\os_{\np,\nm}(\hat x,\hat y)\sim
 (\pi\, \np!\, \nm!)^{-1/2}  \, (\hat x+i\hat y)^{\np} \, (\hat x-i\hat y)^{\nm}
 \, e^{-(\hat n^2+\hat n^2)/2} \,,
\end{eqnarray}  
obtained by replacing $\partial_{\hat x}$ by $-\hat x$, $\,\partial_{\hat y}$ by $-\hat y$, and a similar expression for $\os_{\NP,\NM}(\hat X,\hat Y)$. One obtains
\be  \label{Psia} 
\Psi_{n,\Mv} 
\sim n^{-n-1} \,  (\pi\, \np! \,\nm! \, \NP! \,\NM!)^{-1/2}
\, u^{\np} \, \bar u^{\nm} \, \hat R^{\NP+\NM} \,.
\ee
Recalling that $\np+\NM=\nm+\NP=n-1=l$ and comparing (\ref{binom}) with (\ref{Psia}) 
we obtain (\ref{decomp-n}-\ref{cM}).

\section{Appendix B. Calculation of the asymptotic tunneling wave function and of the width}

We start from (\ref{Gamow-APP}-\ref{K}). Neglecting the ${(1-m^2)/\eta^2}$ term, the roots of $p_\eta^2(\eta)$ (entrance and exit of the tunnel) are 
$\eta_{\rm in}\simeq-2Z_\eta/\en
$, ~$
\eta_{\rm ex}\simeq - 2\en /F+2Z_\eta/\en $. 
They are slightly complex. We choose the determination
\begin{equation}  \label{p analy} 
p_\eta \simeq %
(1/2)\, \sqrt{iF} \  \sqrt{1-\eta_{\rm in}/\eta} \ \sqrt{-i(\eta-\eta_{\rm ex})} \,,
\end{equation}  
which has cuts along the lines $[0, \eta_{\rm in}]$ and $[\eta_{\rm ex}, -i\infty]$. 
In the tunnel region $[\Re\eta_{\rm in},\Re\eta_{\rm ex}]$ it gives $\Im p_\eta(\eta)>0$,
corresponding to an evanescent wave. 
In the {after-tunnel} region $[\Re\eta_{\rm ex},+\infty]$ it gives $\Re p_\eta>0$, corresponding to an outgoing wave. 
We assume
$\Re\eta_{\rm in} \ll \eta_0 \ll \Re\eta_{\rm ex}\ll \eta$ and $|S(\eta_{\rm in},\eta_{\rm ex}|\gg1$.
The integration contour in (\ref{K}) must avoid crossing the cuts of $p_\eta$, therefore pass \emph{above} $\eta_{\rm ex}$ in the complex plane. One obtains 
\begin{equation} 
  \label{KetaT} 
\begin{aligned}                                       
S(\eta_0,\eta)  \simeq & 
\sqrt{{-\en/2}}  
\left(\eta_0+\frac{4 \en}{3F} \right) 
+ ({\eeta/2}) \ln \left(\frac{-8\en }{F \eta_0 } \right) 
\\
& + {(i/3)} \, \sqrt{F} \, \left(\eta+2\en/F\right) ^{3/2}
+{i\pi\eeta/2} 
\,.  
\end{aligned}
\end{equation}
The first line is $S(\eta_0,\eta_{\rm ex})$, the second is $S(\eta_{\rm ex},\eta)$.

For $\feta_{\rm S}( \eta _{0}) $ we use $\eta^{-1/2} \feta_{\rm S}( \eta) e^{im\phi} \equiv 
\os_{\NP,\NM}(\hat X,\hat Y)$ and Eq.(\ref{fosca}) (changed with capital letters):
\be \label{aeta} 
\feta(\eta_0)\sim
 (\pi\, \NP!\, \NM!/\nu)^{-1/2}  \, (\eta_0/\nu)^{\eeta/2}
 \, e^{-\eta_0/(2\nu)} \,.
\ee
Using $p_\eta(\eta _{0})\simeq i/(2n)$, $\,p_\eta(\eta) \simeq \sqrt{F\eta}/2$ in (\ref{Gamow-APP}) we arrive at 
\be \label{Thetaa} 
\begin{aligned} 
\feta_{\rm G}(\eta) \sim &
\left(\pi\, \NP!\, \NM!\sqrt{F\eta}\right)^{-1/2}
 \left(\frac{4}{F\nu^3}\right)^{\eeta/2} 
\exp\left\{\frac{-1}{3F\nu^3} \right\}
\\
& \exp\left\{ \frac{i}{3}\sqrt F  \, \left(\eta+2\en/F\right) ^{3/2}
+{i\pi\eeta/2} +{i\pi/4} 
\right\}.
\end{aligned} 
\ee
The ionization rate is the flux through the paraboloid $\eta$= constant $\gg n^2$. In the mixed representation (\ref{mix}), it reads
\be
\gamma = 2\nu p_\eta \,  |C|^2 \, |\feta(\eta) |^2 \int d\hat x\, d\hat y\, | \os_{j,k}(\hat x,\hat y) |^2 \,.
\ee
The last integral equals 1. In $|\feta(\eta) |^2 $ we consider the second line of (\ref{Thetaa}) as a pure phase factor, neglecting $\Im\en$. One obtains Eq.(\ref{Slavjanov'}).

 \bibliographystyle{aipproc}

 \end{document}